\documentclass[aps,superscriptaddress,11pt,final,notitlepage,oneside,onecolumn,showpacs,showkeys,amsmath,amssymb]{revtex4-2}
\usepackage{graphicx}

\begin{document}

\title{Construction of Chiral Cosmological Models Unifying Inflation\\ and Primordial Black Hole Formation}

\author{\firstname{Ekaterina O.}~\surname{Pozdeeva}}
\email{E-mail: pozdeeva@www-hep.sinp.msu.ru}
\affiliation{Skobeltsyn Institute of Nuclear Physics, Lomonosov Moscow State University,\\
Leninskie Gory 1(2), Moscow 119991, Russia}
\author{\firstname{Sergey~Yu.}~\surname{Vernov}}
\email{E-mail: svernov@theory.sinp.msu.ru}
\affiliation{Skobeltsyn Institute of Nuclear Physics, Lomonosov Moscow State University,\\
Leninskie Gory 1(2), Moscow 119991, Russia}

\begin{abstract}
We propose the method for construction of $F(R,\xi)$ gravity model, unifying inflation and primordial black hole formation. The proposed models are based on the Starobinsky $R+R^2$ inflationary model, so, the function $F(R,\xi)$ is a quadratic polynomial of the Ricci scalar $R$.
We show that the potential of the  corresponding two-field chiral cosmological model in the Einstein frame can be always found in terms of the elementary functions. The special choice of the function $F(R,\xi)$ allows us to get such a generalization of the hybrid inflation that can describe both inflation, and the primordial black hole formation.
\end{abstract}

% 04.50.Kd 	Modified theories of gravity
% 98.80.-k Cosmology
% 98.80.Cq Particle-theory and field-theory models of the early Universe (including cosmic pancakes, cosmic strings, chaotic phenomena, inflationary universe, etc.)

\pacs{98.80.-k, 98.80.Cq, 04.50.Kd}
\keywords{Modified gravity, primordial black hole, inflation}

\maketitle

\section{Introduction}

A black hole is called primordial if it was formed before the matter dominance epoch of the Universe evolution. The rapidly growing number of direct and indirect observations of black holes with masses beyond the astrophysical range, the occurrence of which is not described by models of stellar collapse, confirms an assumption about the existence of primordial black holes (PBH). In particular, the characteristics of black holes, as a result of the collision of which the detected gravitational waves are generated, are not consistent with the standard concept of black holes formation. The study of the discovered black holes in the centers of galaxies has led to the assumption that it is not a black hole formed due to the accretion of galactic matter, but a galaxy formed around a previously formed black hole~\cite{DolgovUFN}. The hypothesis that a part of the dark matter consists of primordial black holes (PBH) has been proposed in Ref.~\cite{Dolgov:1992pu} (see also Refs.~\cite{Ivanov:1994pa,DolgovUFN,Carr:2020xqk}). The possibility of PBH formation in the early Universe was shown in Refs.~\cite{Ivanov:1994pa,Starobinsky:1992ts} (see  Refs.~\cite{Ketov:2021fww,Ozsoy:2023ryl} as reviews). Large peaks in the amplitude of perturbations during inflation can lead to PBH formation in early post-inflation stage of the Universe evolution.

The Starobinsky $R+R^2$ gravity model of inflation~\cite{Starobinsky:1980te} is the oldest inflationary model that is in excellent agreement with current observation data~\cite{Galloni:2022mok,BICEP:2021xfz,Planck:2018jri}.
This model is not suitable for describing of the PBH formation. There are two ways to obtain the models unifying inflation and PBH formation from the Starobinsky inflationary model. One can either use another function of the Ricci scalar $R$ instead of  $R+R^2$, or add a scalar field~\cite{Pi:2017gih,Gundhi:2020kzm,Cheong:2022gfc}. We know several extensions of the Starobinsky model of inflation, for which the inflaton scalar potential in the Einstein frame has the explicit dependence upon fields and parameters in terms of elementary functions~\cite{Ivanov:2021chn,Pozdeeva:2022lcj}, but these models are not suitable for describing of the PBH formation. It is possible to get such analytical functions $F(R)$ that the inflationary model describes PBH formation~\cite{Saburov:2023buy}, but the corresponding potentials in the Einstein frame cannot be presented in terms of elementary functions.

In this paper, we use the second way and propose the $F(R,\xi)$ gravity model, where $\xi$ is a scalar field. The function  $F(R,\xi)$ is a quadratic polynomial of $R$, so the corresponding potential in the Einstein frame can be found in terms of the elementary functions. We demonstrate explicitly how the potential of the inflationary model suitable for PBH formation can be constructed on the example of a modification of the hybrid inflation potential similar to proposed in Ref.~\cite{Braglia:2022phb}.

\section{$F(R,\xi)$ Models and the Corresponding Chiral Cosmological Models}

The modified gravity model, described by the following action:
 \begin{equation}
S_{R}=\int d^4x\sqrt{-\tilde{g}}\left[F(\tilde{R},\xi)-\frac12\tilde{g}^{\mu\nu}\partial_{\mu}\xi\partial_{\nu}\xi \right],
\label{FR}
\end{equation}
The model \eqref{FR} is equivalent to following two-field model
\begin{equation}
\label{SJFR}
S_{J}=\int d^4x\sqrt{-\tilde{g}}\left[F_{,\sigma}\tilde{R}+\left(F-F'_{,\sigma}\sigma\right)-\frac12\tilde{g}^{\mu\nu}\partial_{\mu}\xi\partial_{\nu}\xi \right],
\end{equation}
where $F_{,\sigma}=\frac{dF}{d\sigma}$,  $\xi$ is a scalar field.

Using the conformal transformation of the metric $g_{\mu\nu}=\frac{2F_{,\sigma}}{M_{Pl}^2}\tilde{g}_{\mu\nu}$, we obtain a chiral cosmological model (see~\cite{Chervon:2019nwq} and references therein) with two scalar fields, described by the following action
\begin{equation*}
\label{FRSE}
S_{E}=\int d^4x\sqrt{-g}\left[\frac{M_\mathrm{Pl}^2}{2}R-\frac{g^{\mu\nu}}{2}\partial_\mu\phi\partial_\nu\phi-\frac{M_\mathrm{Pl}^2}{4F_{,\sigma}}{g^{\mu\nu}}\partial_\mu\xi\partial_\nu\xi-V_E\right],
\end{equation*}
where $\phi$, $\xi$  are scalar fields and
\begin{equation}\label{V_E}
V_E= \frac{M_\mathrm{Pl}^4}{4F_{,\sigma}^2}\left(F_{,\sigma}\sigma-F\right),\qquad     \phi=\sqrt{\frac32}\,M_\mathrm{Pl}\ln\left(\frac{2F_{,\sigma}}{M_\mathrm{Pl}^2}\right).
\end{equation}

To get the potential $V_E(\phi,\xi)$ in the explicit form we should find the function $\sigma(\phi,\xi)$. We assume that $F(\sigma,\xi)$ has the following form:
\begin{equation}\label{Fmodel}
   F(\sigma,\xi)=\frac{M_\mathrm{Pl}^2}{2}\left[X_0(\xi)F_{\mathrm{Star.}}(\sigma)+X_1(\xi)\sigma -U(\xi)\right],\quad\mbox{where}\quad F_{\mathrm{Star.}}=\sigma+\frac{\sigma^2}{6m^2}\,,
 \end{equation}
$X_0(\xi)$, $X_1(\xi)$ and $U(\xi)$ are differentiable functions, $m$ is the parameter. In this case,
\begin{equation}\label{sigma}
    \sigma={}-{\frac {3\,{m}^{2} \left[ \left(X_0(\xi) +X_1(\xi)\right)y-1 \right] }{X_0(\xi) y}},\quad\mbox{where}\quad y=\exp\left(-\sqrt{\frac{2}{3}}\frac{\phi}{M_{Pl}}\right)\,.
\end{equation}

Substituting \eqref{sigma} into \eqref{Fmodel}, we get
\begin{equation}
\label{Fstar(ychi)}
   F_{\mathrm{Star.}}(\sigma(y,\xi))=\frac{3\,m^2\left(X_1\left( \xi \right)y-1\right)^2}{2X_0(\xi)y^2}-\frac{3m^2}{2}X_0\left(\xi \right)\,.
\end{equation}
The corresponding expression for potential is:
\begin{equation}
\label{VEy}
\begin{split}
    V_E&=\frac{3M_\mathrm{Pl}^2{m}^{2}\left(X_0\left(\xi\right)\sigma^{2}+6{m}^{2}U\left(\xi \right)  \right)
}{4\left(3{m}^{2}\,X_0\left( \xi \right) +3{m}^{2}\,X_1\left( \xi \right) +X_0\left( \xi \right) \sigma\right)^2}\\
 &=\frac{M_\mathrm{Pl}^2}{4X_0(\xi)}\,\left[3{m}^{2}{y}^{2}X_0^{2}(\xi) +3\,{m}^{2}\left(X_1(\xi) y-1\right)^{2}+2yX_0(\xi)\left(3\,{m}^{2}{y}X_1(\xi)-3\,{m}^{2}+yU(\xi)\right)\right]\,.
    \end{split}
\end{equation}

So, the choice of the functions $X_0(\xi)$, $X_1(\xi)$, and $U(\xi)$ allows us to get a suitable potential $V_E(y,\xi)$ in the explicit form.
By this way, we can construct the two-field potentials for inflationary models with PBH formation. During inflation the both fields play a role of the inflaton: $\phi$ at the beginning and $\xi$ at the end of inflation.
The investigations of inflationary models with two stages of inflation shows that density perturbations at the time corresponding to the phase transition between the two inflationary stages can be large. The resulting density inhomogeneities lead to a production of black holes~\cite{Pi:2017gih,Gundhi:2020kzm,Garcia-Bellido:1996mdl}.
The potentials in the Einstein frame have saddle points.

Differentiating $V_E$ over $\phi$, we get the extremal point at
\begin{equation}
y_{extr}=\frac{3\,m^2\left(X_0(\xi)+X_1(\xi)\right)}{3\,m^2\left(X_0(\xi)+X_1(\xi)\right)^2+2X_0(\xi)U(\xi)}.
\end{equation}

The condition $\frac{dV_E}{d\xi}=0$ at $y=y_{extr}$ is equivalent to the following differentiation equation:
\begin{equation}
\left(X_0+X_1\right)^{2}{\frac{{\rm d} U}{{\rm d}\xi}}-2\,U\left(X_0+X_1\right){\frac {{\rm d}X_0}{{\rm d}\xi}} -2\,U\left(X_0+X_1\right)\frac{{\rm d}X_1}{{\rm d}\xi}=\frac{2\,U^{2}}{3\,m^2}\,\frac{{\rm d} X_0}{{\rm d}\xi}.
\end{equation}

\section{Modification of the Hybrid Inflation}

Using the proposed method, we can construct the potential of the inflationary two-field model suitable for the PBH formation. 
For example, let us choose the potential in the Einstein frame as follows:
\begin{equation}\label{VEfromR2}
V=\frac{\lambda}{4}\left(\xi^2-\xi_0^2\right)^2+\left(C_0+C_1\xi^2\right)\left(1-y\right)^2+d\,\xi,
\end{equation}
where  $\lambda$, ${\xi_0}$, $C_0$, $C_1$, and  $d$ are constants. This modification of the hybrid inflation~\cite{Linde:1993cn} with an additional linear term has been proposed in Ref.~\cite{Braglia:2022phb}.

It is easy to check, that the potential (\ref{VEfromR2}) corresponds to the following functions:
\begin{eqnarray}
 \label{vjj} U&=&{\frac { 2\left( \lambda\left({\xi}^{2}-{\xi}^{2}_0\right)^2+4\,d\,\xi \right)  \left( {C_1}\,{\xi}^{2}+{C_0} \right) }{M_\mathrm{Pl}^2\left( \lambda\left({\xi}^{2}
 -\xi_0^{2}\right)^2+4\,{C_1}\,{\xi}^{2}+4\,
d\,\xi+4\,{C_0} \right)}}\,,\\
   \label{X0}X_0&=&{\frac {3\,{m}^{2}M_\mathrm{Pl}^2}{\lambda\left({\xi}^{2}-\xi_0^{2}\right)^2+4\,{C_1}\,{\xi}^{2}+4\,d\,\xi+4\,{C_0}}}\,, \\
  \label{X1} X_1&=&\frac {4\,C_1\,{\xi}^{2}+4\,C_0-3\,{m}^{2}M_\mathrm{Pl}^2}{\lambda\left(\xi^2-\xi_0^2\right)^2+4\,{C_1}{\xi}^{2}+4\,d\xi+4\,{C_0}}\,,
\end{eqnarray}
and we get the following chiral cosmological model in the Einstein frame
\begin{equation}
\label{FRSEtheta}
S_{E}=\int d^4x\sqrt{-g}\left[\frac{M_\mathrm{Pl}^2}{2}R-\frac{1}{2}g^{\mu\nu}\partial_\mu\phi\partial_\nu\phi-\frac{y}{2}{g^{\mu\nu}}\partial_\mu\xi\partial_\nu\xi
-V\right]\,.
\end{equation}
 During inflation the both fields play a role of the inflaton: $\phi$ at the beginning and $\xi$ at the end of inflation.

\begin{figure}[ht]
	%\centering
	\begin{tabular}{ccc}
	\includegraphics[width=0.4\linewidth]{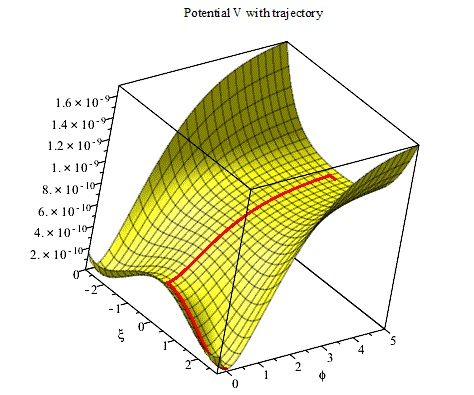}
	\includegraphics[width=0.3\linewidth]{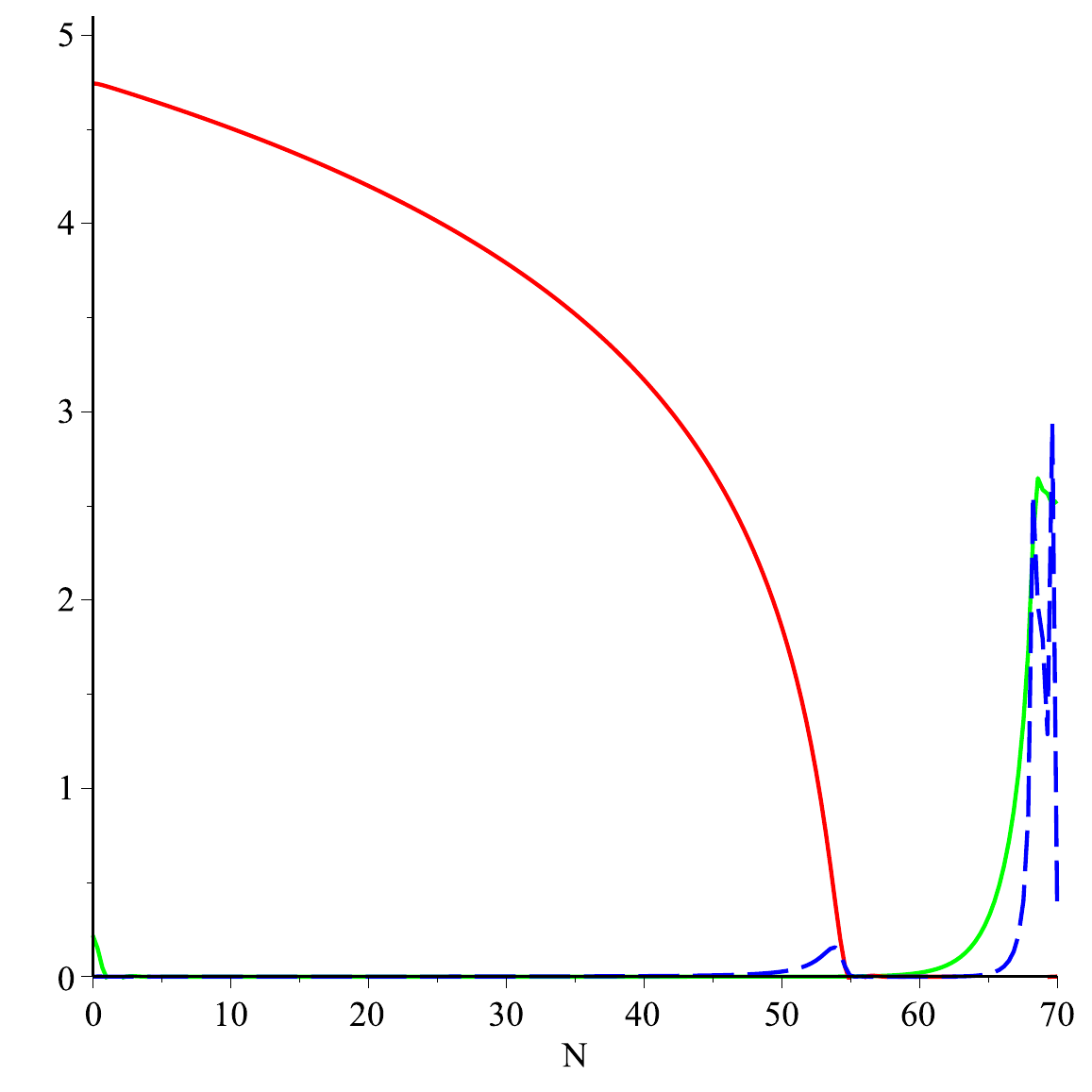}
\includegraphics[width=0.3\linewidth]{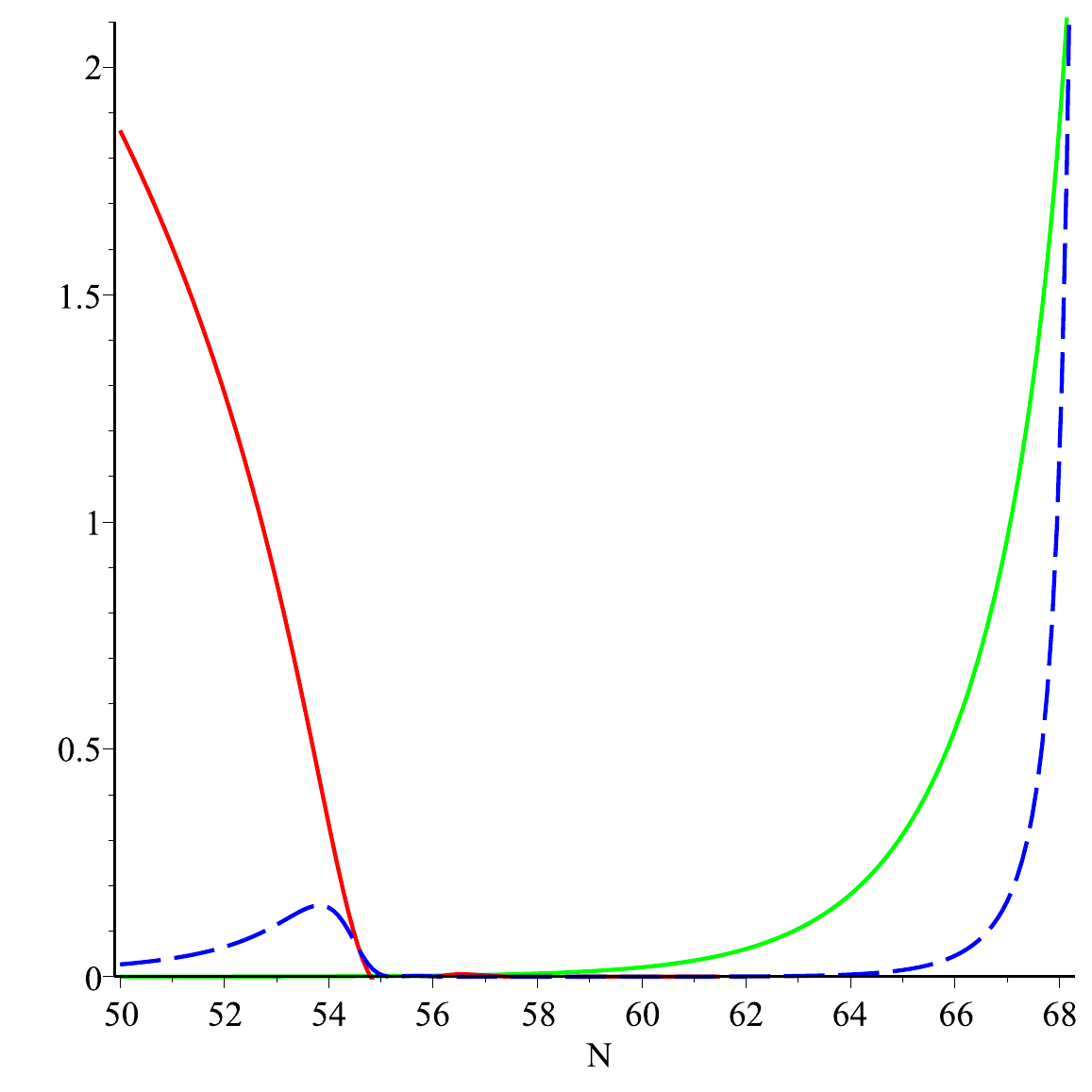}
	\end{tabular}
	\caption{The potential $V(\phi,\xi)$ and the trajectory during inflation (left picture). The fields and the slow-roll parameter $\epsilon_1$ as functions of e-folding number $N$: the blue curve depicts $\epsilon_1(N)$, the red curve corresponds to $\phi(N)/M_\mathrm{Pl}$ and the green curve corresponds to $\xi(N)/M_\mathrm{Pl}$ (center and right pictures). For numerical calculations, the following set of parameters has been used: $m\approx1.47\cdot 10^{-5}\,M_\mathrm{Pl}$, $\lambda\approx3.457\cdot10^{-11}$, $C_0\approx6.483\cdot10^{-10}\,M_\mathrm{Pl}^4$, $C_1\approx1.383\cdot10^{-10}\,M_\mathrm{Pl}^2$, $ d\approx-1.080\cdot10^{-15}\,M_\mathrm{Pl}^3$. }
	\label{Fig1}
\end{figure}

To analyze the obtained model we numerically solve the evolution equations in the spatially flat Friedmann--Lemaitre--Robertson--Walker metric with
$ds^2={}-dt^2+a^2(t)\left(dx_1^2+dx_2^2+dx_3^2\right)$, where $a(t)$ is the scale factor. During inflation, it is suitable to use $N=\ln(a/a_0)$, where $a_0$ is a constant, as a measure of the time.
Inflation is a period of accelerated expansion of the Universe, so the parameter $\epsilon_1=1-\ddot{a}a/\dot{a}^2<1$. Dots denote the time derivative.
In Fig.~\ref{Fig1}, one can see that the potential has a saddle point. The field $\phi$ changes at the beginning of inflation, whereas $\xi$ changes at the end of inflation.

\section{Conclusions}

In this paper, we propose the method to generalize the Starobinsky inflationary model~\cite{Starobinsky:1980te} to $F(R,\xi)$ gravity models, unifying inflation and PBH formation.
Using conformal transformation of the metric, we get chiral cosmological models with two scalar fields. We demonstrate that the corresponding potential in the Einstein frame can be always found in terms of elementary functions. The special choice of the functions of the scalar field $\xi$ allows us
to get a generalization of the hybrid inflation. The potential has a saddle point and looks suitable to describe both inflation and PBH formation.
We plan to present the detail analysis of the proposed model in Ref.~\cite{PV_PBH}.

This study was conducted within the scientific program of the National Center for Physics and Mathematics, section 5 'Particle Physics and Cosmology'. Stage 2023--2025.

%%%%%%%%%%%%%%%%%%%%%%%%%%%%%%%%%%%%%%%%%%%%%%%%%%%%%%%%%%%%%%%%%%%%%%%%%%%%%%%%%%%%%%%%%%%

%%%%%%%%%%%%%%%%%%%%%%%%%%%%%%%%%%%%%%%%%%%%%%%%%%%%%%%%%%%

\end{document}